AI in Design Education at College Level: Educators' Perspectives and Challenges

**Authors:**
Lizhu Zhang1, zhan8909@umn.edu
and Cecilia X. Wang1, ceciw@umn.edu

**Author affiliations:**

1 College of design, University of Minnesota Twin Cities, 1985 Buford Ave., St. Paul, MN 55108, US

**Abstract**

Artificial intelligence has deeply permeated numerous fields, especially the design area which relies on technology as a tool for innovation. This change naturally extends to the field of design education, which is closest to design practice. This has led to further exploration of the impact of AI on college-level education in the design discipline. This study aims to examine how current design educators perceive the role of AI in college-level design education, their perspectives on integrating AI into teaching and research, and their concerns regarding its potential challenges in design education and research. Through qualitative, semi-structured, in-depth interviews with seven faculties in U.S. design colleges, the findings reveal that AI, as a tool and source of information, has become an integral part of design education. AI-derived functionalities are increasingly utilized in design software, and educators are actively incorporating AI as a theoretical framework in their teaching. Educators can guide students in using AI tools, but only if they first acquire a strong foundation in basic design principles and skills. This study also indicates the importance of promoting a cooperative relationship between design educators and AI. At the same time, educators express anticipation for advancements in ethical standards, authenticity, and the resolution of copyright issues related to AI.

## 1. Introduction

Artificial intelligence (AI) has revolutionized many industries and deeply influenced daily life in today's rapidly evolving technological landscape. Such as self-driving cars, Apple Vision Pro, and virtual assistants have significantly changed user behaviors and needs. The concept of AI was first introduced at the Dartmouth Conference in 1956 and has progressed significantly over six decades (Russell & Norvig, 2021). With its continuous development, AI has permeated numerous fields. The integration of AI into education has become a topic that is receiving more and more attention from researchers around the world. The number of countries incorporating AI into education rose from 53 during 2000–2004 to 82 during 2005–2009, and further increased by 33% from 94 to 125 countries between 2010–2014 and 2014–2019 (Song, P. & Wang, X., 2020).

AI has not only changed people's daily life but has also brought a significant wave of change to the **design industry**. Researchers are constantly exploring the use of advanced technologies in design, so this has sparked a discussion on the evolution of the designer's role in the age of AI (Celaschi et al., 2024). The emerging AI brings numerous possibilities and opportunities for design and design education, such as AI-driven tools, machine learning, and intelligent interaction. These technologies offer the potential for greater creativity, increased efficiency, and better user experiences in design. It's worth noting that as technology and design software continue to advance, companies such as Adobe continue to release newer versions of their software with AI features, and platforms such as Figma integrate AI features to help designers realize their ideas more efficiently. AI is disrupting the field by somehow reducing the time and labor costs of design. Design education faces unique challenges. As design tools develop and technology advances, these innovative features will continue to challenge design education.

In the context of the rapid development of AI, AI has prompted deep thinking and change in the college education system. Specifically, this study focuses on the perspectives and attitudes of design educators in the face of the impact of AI and how to adapt to this rapidly evolving technological environment. Faculty's preparation methods, design processes, and students' learning modes are likely to undergo profound changes. Therefore, an in-depth understanding of these changes and challenges is essential for the correct and effective utilization of the potential of AI to promote the benign and innovative development of design education and better adapt to the future needs of the design industry.

## 2. LIterature review

### 2.1Ai in Education

During the past few years, research on the impact of AI on college-level education has grown significantly, reflecting its influence across various disciplines. Scholars from fields such as education (e.g., Al-Zahrani et al., 2023; Chu et al., 2022), computer science (Dai & Ke, 2022), psychology (Pineda et al., 2010), and ethics (Zawacki-Richter et al., 2019) have explored diverse aspects of AI implementation in college education, highlighting its importance and potential. AI has created a transformative power in today's education that revolutionizes teaching methods and significantly improves the efficiency and effectiveness of teaching and learning. AI profoundly impacts student learning by improving skills, facilitating collaborative learning, and fostering productive learning environments (Kuleto et al., 2021).

AI has gained widespread popularity in education as an innovative tool due to its unique capabilities(Almaiah et al., 2022). Huang's study also emphasizes the positive implications of AI being introduced into teaching and learning structures in college-level education. Emphasizing its ability to improve the quality of teaching and learning experience(Huang, 2018). AI-driven systems, such as intelligent tutoring systems, personalized learning platforms, and automated grading systems, are redefining interactions between educators and students.

These AI-powered systems provide students with personalized and flexible learning experiences. Allowing educators to gain meaningful insights from student behavior and needs to improve teaching strategies. As highlighted in the report "The Future of Artificial Intelligence in Teaching and Learning 2024", AI delivers educational content that meets the personalized learning needs of students, and providing customized learning experiences leads to improved engagement and educational outcomes.

### 2.2 Ai in Design Education

AI is reshaping design education through innovative methods and enhanced efficiency. From auto-generative design software to computer vision to natural language processing tools, AI is revolutionizing existing design processes. GE and FAN's research examined how AI supports rapid prototyping and the resolution of complex design problems, significantly impacting both design education and practice (Ge & Fan, 2014). Yamen investigated the ability of AI to simulate user behavior, enabling students to refine their designs iteratively and understand usability from the user's perspective (Yamen Idelbi,2024). Adaptive AI technology allows for individualized task design based on each student's skill level, which improves learners' creative and critical thinking skills.

AI is not only a tool for education. It has also expanded beyond academic settings into the design industry, influencing its practices. As a result, design education needs to be repositioned, and students need to be equipped with the skills required for future jobs, especially in user-centered collaborative design. (Matthews et al.,2023)  Their findings also emphasize that current design education is behind in responding to the demands of the automation age and that further research and reforms are necessary to effectively bridge this gap.

2.3 Challenges of AI in Design Education

Although the extant research examines the potential and challenges of AI in design education, the direct experience and contribution of active design educators are still unexplored. Al-Zahrani et al. highlight the importance of understanding how educators perceive the roles and limitations of AI and their strategies for coping with technological change in teaching. They point out that bringing in the reflections of educators with first-hand experience can help fill this research gap (Al-Zahrani et al. , 2024).

To address this research gap, this study focuses on the following key questions: How do design educators define AI's role in college-level design education? What are their primary concerns about integrating AI into college-level design teaching? By investigating these questions, this study aims to provide insights to effectively address the challenges posed by AI in design education, to ensure the effective integration of AI into educational practices and to provide positive implications for design education.

3.Methods
3.1 Overview
The purpose of this research is to analyze the perspective of educators regarding AI in college-level design education. Taking into consideration that AI has been steadily getting involved in all fields, the use of AI in design education has also become critical due to its immense influence on designers and design tools. To explore this phenomenon, researchers will conduct interviews with seven active college design faculty members to gain insights into the benefits and challenges AI presents in college-level design education.

3.2 Study context

Since the research topic is about analyzing educators' perceptions of AI in college-level design education, this study is based on current faculty of design colleges. Considering the validity and economy of the research, the researcher chose design college faculty in the U.S. as the research subjects and used online interview research for data collection. In this way, researchers can better understand the feedback in real design education and thus better explore the role and impact of AI in design teaching.

3.3 Data collection

Specific data collection methods

Regarding the data collection, researchers performed online one-on-one interviews, which were then recorded in a video format for the analytical purposes. The study population included seven current faculty members from design colleges, purposefully selected to represent a diverse range of ages, genders, teaching experiences, and educational backgrounds. This approach facilitated a deeper understanding of faculties' perspectives and perceptions regarding the integration of AI in design teaching. The researchers have submitted this study for IRB review and it has been determined as not human research.

Interview themes

Interviews were conducted with current faculty members at design colleges based in the United States. The study participants were individual researchers, and these interviews were anonymized; audio recording sessions were de-identified and transcribed verbatim with the consent of the participants. All participants had experience with AI and utilized it to varying degrees in their daily teaching.

The interviews lasted approximately 45 minutes. These interviews were conducted utilizing Zoom's video conferencing one-on-one format and were discussed separately. 1. What courses are you taught in your design college? 2. What AI technologies do you use to teach design? 3. Could you share your personal experience with AI in design education? How can AI help you with your design teaching work? Did you teach AI or AI features to students? 4. What challenges do you face with AI in your teaching? 5. How do students respond when you use AI to teach design? 6. Would you encourage students to use AI in design? Why? 7. How does AI impact your teaching and design research?

Developed the interview questions

The first step was to understand the personal information of the respondents, including basic personal information, age, and gender, and to understand the basic information, such as the position in the school and the courses currently taught. Secondly, researchers tried to ask the interviewees to share their experiences about using AI, which included the experience of teaching AI functions and utilizing AI to prepare or teach classes. After several pilot interviews to summarize the experience, the interview questions were adjusted in order to avoid the state where the pre-interview questions were extended, leading to later questions that had already been answered. Then, try to guide the interviewees; for example, if there are some questions that the interviewees do not have specific examples, refer them to speak their minds, or try to know the reason why there are no particular examples at present.

After obtaining consent from the participants, an online video format was used to conduct all interviews, which were audio-recorded and transcribed with the consent of the interviewees. These transcripts will be used for data analysis and preparation of the study report.

### 3.4 Participants

3.4.1 Eligibility Crieria：

For this study, respondents were asked to have several requirements: 1) Comfortable speaking English; 2) Experience teaching design and currently working as an instructor and teaching design courses in a design college; and 3) experience using artificial intelligence and applying it to design or teaching. These requirements are intended to ensure that the respondents have some level of understanding of AI and can provide insights and feedback. Researchers have tried to find suitable respondents who meet these requirements and ensure that their privacy and rights are respected and protected.

### 3.4.2 Recruitment strategies

In order to find interviewees who met the requirements of the study, researchers used two recruitment methods: first, the primary researcher as a Ph.D. student in the College of Design, invited faculty members in the college who were interested in AI and who used AI in their teaching to conduct research interviews. In addition, researchers posted an online recruitment for the research project for faculty members in the College of Design, inviting qualified faculty members to conduct interviews. Using these two recruitment methods aims to broaden the range of interviewees so that researchers can gain more information and insights and explore the use of AI in design education from different perspectives. Researchers will ensure that the rights and privacy of all interviewees are fully protected.

### 3.4.3 Demographic information

| Name | Age | Gender | Teaching year | Course teaching |
|------|-----|--------|---------------|-----------------|
| A | 35-40 | Female | 10 | UX, interactive design |
| B | 50-55 | Female | 23 | UX, Service design |
| C | 40-45 | Male | 12 | Graphic design, Typography |
| D | 30-35 | Female | 4 | Graphic design |
| E | 40-45 | Male | 7 | Package design |
| F | 30-35 | Male | 6 | Product design |
| G | 35-40 | Female | 9 | Graphic design, color |

The study population comprised seven practicing design faculty members from universities in the U.S., including four females and three males, with an average teaching experience of ten years. These educators come from diverse design backgrounds: some emphasize academia, others focus on creative expression, and some center their work on practical design, managing their design studios. This varied sample selection captures a broad spectrum of perspectives from faculty members at different career stages and with distinct professional focuses regarding the integration of AI in design teaching. Through these interviews, the researcher can understand the impact of AI in college design teaching contexts from educators' perspectives, providing valuable insights for future design education.

### 3.4 Data Analysis

Data analysis seven interview recordings were transcribed automatically after the interviews by Zoom, coded, and analyzed on NVivo software using thematic analysis (Braun and Clarke2006). The six steps proposed by Braun and Clarke were followed: (1) familiarization with the data; (2) initial coding; (3)

identifying themes; (4) reviewing the themes; (5) defining and naming themes, and (6) report writing. Only semantic themes were identified.

Data collected from seven participants were categorized according to which questions the participants responded. Data categorization also included repeated readings of the transcribed data, labeling them according to different questions to better understand each response, and documenting the consistency of responses across participants. Additionally, considering the diverse backgrounds of the participants, the researcher was able to analyze the cases provided by the respondents further by incorporating the backgrounds of the interviewed instructors teaching different courses and further analyzing the cases provided by the respondents.

Based on the research questions, affinity diagramming was prepared and set through in this study (Haskins, et al., 2020). The researcher wrote each emerging code on a separate sticky note and identified the codes that overlapped with others. Overlapping codes were aggregated based on their commonalities until they were mutually exclusive. After categorizing the codes by their relationship, including similarity, differences as well as hierarchy, three levels of themes (i.e., high-level, mid-level, low-level) from this iterative process resulting in overall selection themes.

### 4.Result
4.1. Educators Define AI's Role in College-level Design Education

The continuous development of AI has introduced it into daily design work. For instance, mainstream design software such as Adobe Photoshop, Illustrator, After Effects, and Firefly have also been integrated with AI functions individually. AI functions are also widely used to improve design efficiency and promote design innovation. This technological progress has introduced new challenges for design education, prompting educators to reconsider how to integrate AI functionalities into their curricula effectively.

> *"We've observed that with the explosion of AI, many design software tools incorporate AI functionalities. These features drastically reduce the time designers spend on production tasks. So, when I teach undergraduate design software courses, I plan to integrate these AI features into the curriculum to help students realize their design ideas more efficiently and effectively."*
> （Participant B）

AI is not only integrated into design software but has also expanded into teaching and research through generative models like Midjourney. As a text-to-image AI tool, Midjourney aids students in exploring the distinctions between AI-generated and human-created works.

> *"I've used Midjourney to generate AI images and compare them with human-created images to demonstrate the differences. This helps students better understand how AI works and their unique characteristics."* （Participant D）

In class practices, AI could offer to offer diverse support to students in the design and research process. For example, students use tools such as ChatGPT to provide an initial framework and inspiration for design solutions.

> *"I recommend using AI as a tool. For instance, in my UX class, students use ChatGPT to construct user journeys and explore what it might suggest and how they could solve design challenges."* （Participant A）

Moreover, many participants agreed that AI serves as an effective auxiliary tool in teaching and research, supporting information gathering, data organization, knowledge expansion, and lesson preparation.

> *"I use ChatGPT to prepare lectures, such as collecting resources and case studies of renowned product designers to enrich the course content."* （Participant G） AI can improve efficiency in some situations. *"Using Midjourney, our research team completed character design in two days. A process that traditionally took two months, significantly improving the research efficiency."* （Participant A）

While AI is an effective tool for improving efficiency and enriching educational practices. However, it is important to ensure that AI does not overshadow students' ability to innovate and to teach them how to work creatively with AI.

> *"I guide students to explore the integration of AI into UX design research by comparing traditional methods, AI methods, and hybrid approaches."* （Participant E） *"The key is teaching students to collaborate with emerging technologies like AI through design, rather than simply learning the technologies themselves."* （Participant C）

The wide application of AI provides strong support for design education. By integrating it into design software and generation tools, efficiency has been enhanced, teaching practice has been enriched, and support has been provided for the research process. However, educators highlight that while taking innovation as a core skill, guiding students to collaborate creatively with artificial intelligence is necessary.

4.2. Educators' attitudes and recommendations on students' use of AI in design learning

| Positive Feedbacks | Negative Feedbacks |
|---|---|
| **Personal perspective：**<br>*"Prefer to use AI."*(Participant A)<br>*"Incorporate AI in design principles and theory"*(Participant A)<br>*"AI can take answers for you."*(Participant C)<br>*"AI really makes your life easier."*(Participant D) | **Trustiness：**<br>*"Trustiness is a really key issue"*(Participant A)<br>*"AI doesn't really help us, do you fully trust the results it generates?"*(Participant F)<br>*"Fake news and like wrong information"*(Participant G)<br>*"I always double-check the AI result"*(Participant A) |
| *"AI as my personal editor so that's very helpful of it."*(Participant E) | *"Need to confirm the result"*(Participant D) |
| *"AI can help me to manage my time."*(Participant D) | |
| *"AI showed me step by step how to do the new thing."* （Participant E) | **Expect more developments:** |
| *"Accepting AI based designs."*(Participant F) | *"The relationships between AI and human beings are still unstable."*(Participant A) |
| | *"AI is still Developing."*(Participant E) |
| **Encourage students to use it:** | *"AI still could not be able to help you to think."*(Participant A) |
| *"Encourage students to use and understand AI before they start using it in their practice."*(Participant A) | **Student Use Concerns:** |
| *"Encourage students to use，not for beginners."*(Participant D) | *"Sometimes, if they use AI, then there's no point in the students learning a new tool."*(Participant B) |
| *"Encourage them to do it."*(Participant F) | *"Rather than simply learning the technologies themselves."*(Participant C) |
| *"Encourage students to use it appropriately with ethics."*(Participant B) | *"Avoid dependency."*(Participant E) |
| | *"Don't use it to write your paper."*(Participant C) |
| **A resource:** | |
| *"AI can be a resource. Use it to gather research and resources for your paper."*(Participant C) | **Failure to achieve expected results:** |
| *"Really want to keep it as a tool."*(Participant B) | *"Never utilize it to be the end result."*(Participant G) |
| *"Such as collecting resources and case studies."*(Participant G) *"Students can use AI to generate initial concepts."*(Participant E) | *"Not very helpful for use it to create a real design"*(Participant D) |
| *"I use AI a little bit more like a teaching assistant to support."*(Participant F) | *"Sometimes we cannot get the result we want."*(Participant B) |
| *"Incorporating AI more is in their final designs to produce a mock-up."*(Participant E) | *"Fails to accurately meet the requirements, for example, by failing to correctly draw complex object details."*(Participant C) |
| | *"The output is lower standard"*(Participant B) |
| **Effective:** | *"If you start looking closer to the image AI generates you will see errors and flaws."*(Participant F) |
| *"Drastically reduce the time to help students realize their design ideas more efficiently and effectively."* (Participant B) | **Ethical Concerns:** |
| *"Significantly improving the research efficiency."*(Participant A) | *"A gray area, requiring further legal clarification."* (Participant F) |
| | *"Students need to understand the boundaries and rules of AI usage to avoid problems."*(Participant G) |

*Table 1：The table shows participants' positive and negative views on AI in higher university design education.*

All participants emphasized the importance of encouraging students to incorporate AI into design practice and research.(Table 1) AI has permeated every aspect of current life, and preventing its use through educational restrictions is neither realistic nor effective.

*"I see a strong connection to the history of design. Before computers, we used hand-drawn methods for posters and illustrations. We never said, 'Don't ever use computers.' That would*

*have left us in the dark ages. AI is just another tool we need to learn how to collaborate with, as ethical creatives."* （Participant B） *"This emerging technology is a powerhouse. Our students may know more about AI than we do... because this is the world we live in."* （Participant G）

Some participants suggested a more restrictive approach to its use. They agreed that AI should remain a tool focused on its assistive capabilities rather than becoming directly involved in design creation or academic writing, particularly in foundational university courses. It is important for lower-level college design students to gain knowledge of design theory and skills, rather than relying on AI to accomplish design tasks. Students must develop critical thinking skills when using AI to ensure the core values of learning and creativity are preserved.

*"Students can use AI to generate initial concepts, but they must critically refine these outputs to avoid dependency."* （Participant E）

*"I require students to show sketches and handmade processes rather than submitting AI-generated design outputs directly."* （Participant F）

This approach not only safeguards students' hands-on abilities but also helps them understand fundamental design principles and the creative process.

AI can serve as an efficient tool for data and information collecting but must have clearly defined boundaries to prevent undue interference in academic outputs.

*"AI should be like Google—a resource. You record what you find, but you don't use it to write your paper. Instead, you use it to gather research and resources for your paper."* （Participant C） *"Just as computers replaced hand-drawing, AI is a new tool we need to learn to use—provided we maintain ethical awareness."* （Participant D）

The participants expressed that the application of AI in education, and design practices, needs to be strictly as a supportive tool that helps students to learn and practice more, but never to replace their ideas and effort. Educators' intent with helping students use AI in a prudent way is to preserve and stimulate the core values of design education: innovation and critical thinking.

4.3.Concerns About AI Integration in Design Education and Practice

As AI is more and more used in design, educators  and researchers start to focus on its limitations and urgent issues. These dilemmas concern the meeting of the complex requirements of the design, both reliability and transparency of the content produced by the AI, as well as questions of copyright and ethics. While AI provides significant support in various scenarios, its capabilities often fall short when facing intricate demands.

Participant B shared an experiment where she tasked AI with generating an image of a breadstick riding a cucumber as bicycle wheels. Despite multiple attempts, the AI failed to produce a satisfactory result, falling short compared to a composite photo created by students using graphic software methods.

*"In complex design scenarios, the AI often fails to accurately meet the requirements, for example, by failing to correctly draw complex object details." (Participant C) "AI doesn't really help us realize ideas accurately. For example, when automatically generating portraits, AI often fails to correctly draw the details of human fingers." (Participant F)*

The accuracy and credibility of AI-generated content remain critical concerns. Participants emphasized that users must carefully evaluate and verify AI outputs, especially in academic research, where AI-recommended resources and information might be false or misleading.

*"This comes down to trust in AI. Do you fully trust the results it generates? What do you do if the answer is wrong? You may need to rephrase your query and get a different response. For instance, if my students use AI to explore a research area, it might generate a list of papers, but I always ask them to verify the papers' authenticity through Google Scholar." (Participant A)*
*"When students use AI-generated research references, they need to confirm their validity using platforms like Google Scholar." (Participant D)*

Copyright issues associated with AI-generated content have raised further concerns among educators. Questions about ownership whether it belongs to the AI tool developers, the original content providers used for training, or the AI user remain unresolved.

*"The copyright of AI-generated stuff is still a gray area. They require further legal clarification." (Participant F)*

*"Students really need to understand the do's and don'ts when it comes to using AI, so they can steer clear of any issues later on, whether in school or in their careers." (Participant G)*

The participants have also highlighted different concerns regarding the limitations and challenges of AI. Although artificial intelligence has demonstrated efficiency in many design scenarios, it still faces many challenges, including understanding complex creative requirements, producing more reliable results, as well as design copyright and ethical issues. To address these challenges, educators must guide students to use artificial intelligence responsibly to ensure the scientific and compliant application of technology.

## 5 Limitation

This study interviewed seven current faculty members ranging in age from 35 to 55 years old. Educators in this age range tend to be in the middle stages of their careers and typically have a more established approach to teaching. However, missing from the study sample were younger educators and senior educators nearing retirement, two groups whose attitudes and acceptance of AI may differ.

Also, all participants were from Minnesota, and this regional perspective of the study helped to analyze in depth the current state of design education, but the perspectives captured may be shaped by the unique characteristics of the Midwestern collegiate environment. Future research could expand the sample to gain a broader academic perspective.

AI is developing rapidly and the industry's attention to it continues to grow. As AI tools continue to iterate, the findings of this study may be limited by time, and some of the conclusions may need to be revisited as the technology advances.

## 6 Conclusion

This study conducted in-depth interviews with seven current faculty members from design schools in U.S. universities to explore their experiences with and attitudes toward AI in college-level design education and research. It examined the current state of AI in design education, its potential impact on teaching and learning, and educators' perspectives on its opportunities and challenges. The findings show that AI has been integrated into design education, and should serve as an auxiliary tool to enhance efficiency and streamline processes. AI tools like Midjourney and ChatGPT are widely used for course preparation, data collection, and design execution. However, AI also presents limitations in handling complex creative tasks. In particular, the design requires careful attention to detail or a high degree of innovation. It follows that while AI offers useful support, it cannot replace the creativity of students.

Moreover, while AI offers numerous benefits for teaching, educators emphasize that the key objective of design college-level education is to develop students' foundational understanding of design theories, skills, and creative processes. Students should not rely on AI to complete their learning tasks. Especially in foundational courses, priority should be given to mastering core design principles and methods, with AI serving only as a supplementary tool. Cultivating critical thinking and ethical awareness is essential in AI-era design education to ensure students uphold academic and professional responsibility in their technological applications.

This research also highlights the challenges of AI in college-level design education, including the reliability of generated content, the transparency of data sources, and unresolved copyright and ethical concerns. The educators advocate using verification ways to ensure the credibility of AI-generated content. Students need to understand the rules and boundaries of AI usage to mitigate academic risks. At the same time, copyright issues will require further legal clarification and support.

Overall, AI has entered the field of design education and has gradually become an indispensable tool for both educators and students. While it presents great advantages, it also brings up challenges that need to be seriously addressed. When applying AI, faculty should ensure that its use is grounded in enhancing students' foundational skills, fostering critical thinking, and reinforcing ethical awareness. In doing so, AI can strengthen, rather than weaken, the core values of creativity and innovation in design education.